\documentclass[aps,preprint,prb]{revtex4}         
         
\usepackage{amsfonts}  
\usepackage{amsmath}         
\usepackage{amssymb}         
\usepackage{graphicx}         
         
\begin{document}            

\title{Long time relaxation of interacting       
electrons in the regime of hopping conduction}       
               
\author{D. N. Tsigankov$^1$, E. Pazy$^2$, B.D. Laikhtman$^3$, and A. L. Efros$^1$}        
      
\affiliation{$^1$Department of Physics, University of Utah, 
Salt Lake City, Utah 84112\\
$^2$Department of Physics, Ben-Gurion University of the Negev,
Beer-Sheva 84105, Israel\\
$3$Department of Physics, Hebrew University, Jerusalem 91904, Israel }

\begin{abstract}

Using numerical simulations we studied the long time relaxation of 
the hopping conductivity. Even though no modern computation is able to simulate 
the behavior of a large size system over minutes or hours so as to observe the relaxation,
still we have been able to show that the long time relaxation and aging effect 
observed in experiments can be explained in terms of slow transitions between 
different pseudoground states. This was achieved by showing that 
different pseudoground states may have different conductivities and that 
the dispersion of conductivities is in agreement with the experimental data. 
We considered two different two-dimensional models with electron-electron interaction: 
the lattice model and the random site model, corresponding to ``strong'' and ``weak''
effective disorder. For the lattice model, effectively strong disorder, we have shown 
that the universality of the Coulomb gap, which is 
responsible for the universal Efros-Shklovskii law for the conductivity, suppresses the long time relaxation of 
conductivity since the universality strongly decreases the dispersion of 
conductivities of the pseudoground states.    
\end{abstract}
\maketitle       
\section{Introduction}       
The study of electron-electron  interactions in the localized regime         
was initiated by Pollak\cite{Pollak70} and        
Srinivasan\cite{sri}. Later on Efros and Shklovskii(ES) \cite{ES}         
showed that the single particle density of states (DS)         
tends to zero as the energy tends to the Fermi energy. This        
phenomenon, called the Coulomb gap, is due to the long range        
part of the Coulomb interaction  which, in some sense, remains        
non-screened. In fact, the Coulomb gap results from the Coulomb law        
and from the discrete nature of the electron charge. In their first works 
ES claimed that the DS in the Coulomb gap has a 
universal form, depending only on electron charge $e$ and dielectric constant        
$\kappa$. Then DS $\sim |\epsilon| ^{D-1}\kappa^D/e^{2D}$, where $D$, is the 
spatial dimension and $\epsilon$, the single-electron energy  
whose reference point is the chemical potential. 
This expression for the DS is  the only        
combination of the energy and the electron charge which has a proper        
dimensionality. It has been shown later on \cite{pik} that in 2D-case 
the above universality is exact only for strong disorder. 
In the 3D-case the question was never studied in detail,       
however deviations from the quadratic law have been reported\cite{Mobius92}.       
        
Simple quantitative  arguments which assume that the         
single-particle  excitations are responsible for         
the variable range hopping (VRH) lead to        
the so-called  Efros-Shklovskii (ES) law\cite{ES}, which has been observed        
experimentally in many materials        
\begin{equation}        
\sigma_c \sim (\gamma e^2/T)\exp\left(-(T_0/T)^{1/2}\right),        
\label{eq:ES}        
\end{equation}        
where $T$ is the temperature and $T_0=\beta_0 e^2/\kappa a$, $a$ is a localization length of electrons,
$\kappa$ is an effective dielectric constant of the media above and below 2D-gas.
A  self-consistent type of percolation approach (See references        
in\cite{lev}) gives  $\beta_0=6.5$. The hopping length is given by       
$R_C\approx (a/4)(T_0/T)^{1/2}$.        
       
Recently\cite{ourprl}, using the so-called lattice model,
the ES law has been checked in detail by computer simulation.
The Hamiltonian of this model is formulated on a square lattice and has a form       
\begin{equation}        
H = \sum_{i}\phi_{i} n_{i} + \frac{1}{2} \sum_{i \neq j}        
\frac{\left(n_i - \nu \right) \left(n_j - \nu \right)}{r_{ij}},        
\label{eq:ham}        
\end{equation}        
where $n_i = 0, 1$ are occupation numbers.        
The quenched random site energies $\phi_i$ are distributed        
uniformly within the interval $[-1, 1]$, and the average        
occupation number  $\nu$ , is taken to be 1/2. The        
magnitude of the quenched disorder is enough so as to provide the universal        
Coulomb gap at all energies which are  important within the temperature        
range under study\cite{pik}. In what follows the lattice constant is taken to be        
the unit of length. The nearest neighbor Coulomb energy which is in this case equal       
to the amplitude of the disorder is considered both as the energy and the temperature unit.       
    
Simulations of the conductivity in the lattice model\cite{ourprl} confirm the ES law in all details, i.e.,
the pre-exponential factor, T-dependence and $a$-dependence. It has also been        
shown that simultaneous transitions of two electrons do not play any role.       
Arguments have also been given that any many-electron excitations are not important. Therefore 
simultaneous transitions of electrons were not
included in the simulation below. 

Glassy properties, due to both  randomness and the long range Coulomb interaction,
are another interesting manifestation of electron-electron interactions in such a system. Davies, Lee, 
and Rice \cite{daviesprl},\cite{daviesprb} were the first to raise this issue. They coined 
the phrase ``Electron  Glass''  which is still used, sometimes  also referred to as the ``Coulomb Glass''. 
Both these terms stress the relation of the above electron system to a spin glass system.       
       
We believe that, in the same way as in real glasses, the glassy properties in this electron system are due to those states, 
which have  very close total energies but substantially different sets of the occupation numbers       
$n_i$. Such states have been first observed by Baranovskii {\it et al.}\cite{Baranovskii} during 
the first computer simulation of this system and have been called ``pseudoground states'' (PS's). 
These authors found that the PS's have the same universal Coulomb gap and have concluded
that the existence of PS's is not important for the VRH conductivity because transitions between them are  very slow.
They attributed these slow transitions between the PS's as well as their difference from the
ground state to be a result of a certain amount of many-electron transitions.  

Experiments on the glassy properties of such systems, 
conducted around 25 years later, have confirmed this conclusion, but they have also shown that Baranovskii {\it et al.} 
missed an important feature. Since transitions between PS's take a long time and conductivity of these        
states is not exactly the same, they can serve as a basis for memory effects.       
       
Experiments, started by the group of Ovadyahu\cite{zvi1,zvi2} in 1993, definitely show the relation of this system 
to the ordinary glasses. In these experiments the difference       
in the conductivities of the PS's is not larger than 10-12\%. Similar phenomena was observed by the 
group of Goldman\cite{gol} on ultra-thin films of metals        
near the superconductor-insulator transition. Slow relaxation has been demonstrated by 
Don Monroe {\it at al.}\cite{dm} in compensated GaAs.       
       
The properties of the PS's have been studied recently, mostly by computational       
methods \cite{kogan,perez98,menashe}. Perez-Garrido {\it et al.}\cite{perez98} argued  that transitions between PS take a huge       
time, substantially larger than a time available in  any  experiment. Menashe {\it at al.}\cite{menashe} have proposed a 
different method in order as to study the Coulomb Glass. By completely ignoring the 
tunneling term in the transition probability, still keeping the activation probability  
for the electrons, they performed a thermodynamic Monte-Carlo simulation, 
which transforms a non-ergodic system into an ergodic one. Thereby they were able to obtain
all the thermodynamic properties of the ergodic system, since
the electrons were able to move across the system in a single transition, leading to
effective mixing of all PS's.

The above method only permits calculation of the thermodynamic values, since the number of the Monte-Carlo 
steps can not be related to a physical time. Still, these
authors claimed that the metal-insulator transition in the Coulomb 
glass may coincide with the glassy transition, that occurs due to increase        
of the localization length.  Their method \cite{menashe} is partially employed in this paper.        
       
The goal of this paper is to understand the origin of the long-time relaxation of conductivity,        
observed in the experimental papers cited above. To do so we performed Monte-Carlo simulations 
employing: the lattice model Eq.(\ref{eq:ham}) and the random site model described later on, 
for two dimensional systems. In our simulation we  perturb the system
by adding some extra electrons and then trace the relaxation  
of energy and conductivity with time. One should clearly understand that it is impossible to observe the long-time 
relaxation by a direct simulation. The general reason for this is due to the long range of the Coulomb interaction  and
to the fact that in a real sample many transitions 
take place simultaneously while a computer processor is limited to performing them one at a time as well as
updating the resulting changes in the site energies one at a time. 
A few additional processors cannot help much. Therefore, even using the most sophisticated program, 
we may simulate no more than 40 $\mu$sec of the real time in a system of the size of $100\times 100$ lattice sites.       
       
We have shown that during such available times we can reach an apparent saturation of 
conductivity and energy and that saturation of the DS occurs much earlier. We
interpret this ``saturation'' as a saturation within one PS. Thus, it is rather the transition from the fast 
relaxation within one PS to a very slow relaxation to PS's with lower energy, that we can not observe.
       
Simultaneously with the conductivity we studied the relaxation of  energy. We devised an analytical theory for the energy 
relaxation which fits the simulation data fairly well and serves as a 
reference point for our understanding of the short time relaxation.       
       
To study the long time relaxation due to transitions between the different PS's we came up       
with a new idea. We create different PS's by relaxation from states with different initial distributions of 
electrons. Then we study the difference between the saturated values of the conductivities of the different PS's
if these are different, one should expect that a long time relaxation exists and the total change of the conductivity 
should be of the order of this difference. For the lattice model we obtain no such effects.       
The conductivities of different PS are the same in the limits of our accuracy (about 1-2\%).       
       
Since PS's with different energies were observed by Menashe {\it et al.} as well as
in other papers, we arrive at the conclusion that the effect of long-time relaxation of the conductivity is absent due to the 
universality of the Coulomb gap for the lattice model, in the temperature range which we consider.        

We also performed simulations for the random site model. In this model all disorder comes from the random position of sites. 
Our results for these simulations exhibit a difference in the conductivities of the PS's whose value 
is sufficiently large so as to explain the experimental results of Ovadyahu's group.

The difference between these two models in the 3-dimensional case has been discussed in Ref.\cite{discussion}. 
Note that in this case a glassy transition at non-zero temperature has been claimed\cite{Grannan}.             
Xue and Lee \cite{Xue} performed Monte-Carlo simulations employing the 2-dimensional random site model in which the 
disorder comes about only through the random position of sites. 
These authors found evidence for glassy behavior at low temperature but they claimed the absence of the glassy transition 
at non-zero temperature. The same result was obtained for the Ising model\cite{binder}. 
       
The paper is organized as follows. In the next section we study the relaxation        
of the energy and the conductivity after an extra amount of electrons 
were added to the system in the framework of the lattice model. 
In Sec.\ref{sec:conductivity}, employing the 
lattice model, we study the difference in the conductivities of the PS's, obtained through
different relaxation methods. In Sec.\ref{sec:randomsite} the results for the random site model are presented and discussed.       
         
\section{ Relaxation after addition of extra electrons}       
\label{Sec:Relaxation}
In this section we study the relaxation of the energy and conductivity of the system after it has been initially perturbed
by the addition of a few percents of extra electrons. In principle this is the same procedure employed in the experiments 
of Ovadyahu's group but the time in which we are able to observe
the system is very short. We use the lattice model with Hamiltonian given 
by Eq.(\ref{eq:ham}) and initial filling factor $\nu=1/2$. The extra electrons are       
added randomly on empty sites and the background is adjusted such that the system        
remains neutral.       
       
\subsection{Analytical theory of energy relaxation}

To describe the physics of the short time energy relaxation  we devise the following  
analytical theory.  We start by considering a division of the plain with the 2D 
electrons into regions with a linear size R. If
$\delta n$ is an average density of additional electrons, the charge $Q$ of
each region  is of the order of $Q\sim e\sqrt{\delta n R^2}$, because the
average charge is compensated by the background. The extra energy due to 
electron-electron interaction per a region containing excess charge $Q$ is $Q^2/R$. The
number of regions  per area is $L^2/R^2$, where $L$ is the length of the total
system which is square shaped. Thus the extra energy per area due to all
regions of size $R$ is ${\cal E}=e^2\delta n/\kappa R$. The regions with
the smallest $R$ give the largest contribution. However, due to relaxation
they become neutral faster. So, the main contribution at a time $t$ comes from
the regions in which relaxation has not yet ended. In the 2D case the relaxation
goes with the velocity of the order of the conductivity $\sigma_0$ of the
system\cite{dyakon}. Thus, at a time $t$ only the regions with $R>\sigma_0 t$
have an excessive charge. Finally, the energy per area decreases with time as 
${\cal E}\sim e^2\delta n/\kappa\sigma_0 t$.
Analyzing results for the auto-correlation functions and density-density correlation functions
it is not difficult to find the numerical coefficient in this expression. Let
$n'({\bf{r}})$ stand for the density of additional electrons with their
homogeneous background so that the average value $<n'>$ is zero. The linearized
equations have the form
\begin{equation}
{\bf j}'  = \sigma_{0}{\bf E}
\label{leq1}
\end{equation}
and
\begin{equation}
 e \ \frac{\partial n'}{\partial t} +
    \sigma_{0}\nabla {\bf E} = 0.
\label{leq2}
\end{equation}
One can see that the diffusion current, omitted here, is more important on the
earlier stages of relaxation than the Ohmic current.

In the 3D case the field can be eliminated from these equations resulting in an
equation for $n'$ only. In the 2D case the situation is different and the problem
can be solved using the Fourier transformation,
   \begin{equation}
n' = \frac{1}{L} \sum_{\bf{q}} n_{\bf{q}} e^{i{\bf qr}} \ ,
    \hspace{1cm}
    \phi = \frac{1}{L}
    \sum_{\bf{q}} \phi_{\bf{q}}(z) e^{i{\bf qr}} \ .
\label{eq:us.sf.3}
\end{equation}
Here ${\bf r}$ and ${\bf q}$ are two dimensional vectors in the plane of the 
electrons, $z$ is the coordinate in the perpendicular direction and $\phi$ is the scalar potential.  The equation
for the potential has the form

\begin{equation}
\frac{d^{2}\phi_{\bf{q}}}{ dz^{2}} - q^{2}\phi_{\bf{q}} =
    - \frac{4\pi e}{\kappa} \ n_{\bf{q}} \delta(z) \ .
\label{eq:us.sf.4}
\end{equation}
It can be solved in the regions $z>0$ and $z<0$. Under the condition of
continuity
of $\phi_{\bf{q}}$ at $z=0$ and zero at $z\rightarrow\pm\infty$ the
 solution
is
\begin{equation}
\phi_{\bf{q}} = A_{\bf{q}} e^{- q|z|} \ .
\label{eq:us.sf.5}
\end{equation}
The integration of Eq.(\ref{eq:us.sf.4}) over an infinitesimally
 small interval
around $z=0$ leads to the boundary condition
\begin{equation}
\left.\frac{d\phi_{\bf{q}}}{dz}\right|_{z=+0} -
    \left.\frac{d\phi_{\bf{q}}}{dz}\right|_{z=-0} =
 - \frac{4\pi e}{\kappa} \ n_{\bf{q}} \ .
\label{eq:us.sf.6}
\end{equation}
This condition gives
\begin{equation}
A_{\bf{q}} = \frac{2\pi e}{\kappa q} \ n_{\bf{q}} \ .
\label{eq:us.sf.7}
\end{equation}
Eqs.(\ref{eq:us.sf.5}) and (\ref{eq:us.sf.7}) give
\begin{equation}
{\bf E}_{\bf{q}} = - i{\bf q} \
    \frac{2\pi e}{\kappa q} \ n_{\bf{q}} \ , \hspace{1cm}
    ({\bf \nabla  E})_{\bf{q}} = i{\bf qE}_{\bf{q}} =
     \frac{2\pi eq}{\kappa} \ n_{\bf{q}} \ .
\label{eq:us.sf.8}
\end{equation}
Now the Fourier transformation of Eq.(\ref{leq2}) is
\begin{equation}
\frac{dn_{\bf{q}}}{dt} + \frac{n_{\bf{q}}}{\tau_{\bf{q}}} = 0 \ .
\label{eq:us.sf.9}
\end{equation}
where
\begin{equation}
\frac{1}{\tau_{q}} = \frac{2\pi q}{\kappa} \ \sigma_{0} \ .
\label{eq:us.sf.10}
\end{equation}
This leads to
\begin{equation}
n_{\bf{q}}(t) = n_{\bf{q}}(0) e^{-t/\tau_{q}} \ .
\label{eq:us.sf.11}
\end{equation}
The energy of the fluctuations is
\begin{eqnarray}
{\cal E} & = &
    \frac{1}{2} \int \phi({\bf r},z) e n({\bf r}) \delta(z) d^2 r dz =
    \frac{\pi e^{2}}{\kappa} \
    \int \frac{|n_{\bf{q}}(0)|^{2}}{q} \ e^{-2t/\tau_{q}} \
    \frac{d^2{\bf{q}}}{(2\pi)^{2}} \ .
\label{eq:us.sf.12}
\end{eqnarray}

If the initial state corresponds to randomly distributed electrons,
\begin{equation}
\langle n'({\bf r}, 0)n'({\bf r}',0)\rangle =
    \overline{\Delta n} \ \delta({\bf r} - {\bf r}') \ ,
\label{eq:us.sf.13}
\end{equation}
then
\begin{eqnarray}
\langle|n_{\bf{q}}(0)|^{2}\rangle & = &
    \frac{1}{L^{2}} \left\langle
    \int e^{i\bf{qr}} n'({\bf r},0) d{\bf r}
    \int e^{-i{\bf qr}'} n'({\bf r}',0) d{\bf r}'
                \right\rangle
\nonumber \\ & = &
    \frac{\bar{\Delta n}}{L^{2}}
    \int e^{i{\bf q}({\bf r}-{\bf r}')}
    \delta({\bf r} - {\bf r}') d{\bf r} d{\bf r}' =
    \overline{\Delta n} \ .
\label{eq:us.sf.14}
\end{eqnarray}
The substitution of Eq.(\ref{eq:us.sf.14}) in Eq.(\ref{eq:us.sf.12}) gives
\begin{eqnarray}
{\cal E} & = &
      \frac{\pi e^{2}\overline{\Delta n}}{\kappa} \
    \int \frac{1}{q} \ e^{-2t/\tau_{q}} \
    \frac{d^2{\bf{q}}}{(2\pi)^{2}} =
    \frac{e^{2}\overline{\Delta n}}{8\pi\sigma_{0}t}
\label{exact}
\end{eqnarray}
This result gives the value of the  numerical
coefficient for the above estimate. One should
understand that Eq.(\ref{exact}) is valid at $t<L/\sigma_0$. At such times the energy
relaxation has finished due to the finite system size. It is easy to take the
size effect into account qualitatively. The result is that energy as a function
of $t^{-1}$ saturates at $t\approx L/\sigma_0$. It is important also that the
system is assumed to be ergodic, i.e., it does not contain different PS's with slow
relaxation between them.
       
\subsection{Simulation results for short time relaxation of energy and conductivity}       
All the time-dependent simulation results in this paper are obtained       
by our modification of the kinetic       
Monte- Carlo method  that is presented in the Appendix. For the lattice model our        
computer units (CU) are as follows: the length unit is taken to be the lattice constant $a_0$,
the units of energy and temperature are given by the Coulomb interaction at the lattice constant $e^2/a_0$, 
the time unit is chosen as the reciprocal transition rate due to phonons $\gamma^{-1}$ see Eq.(\ref{eq:ES}),
and the unit of the 2-dimensional conductivity  is given by $a_0\gamma$.         
The numerical values for the time given below, are calculated        
using the assumption that $\gamma=10^{12}s^{-1}$.       
       
The simulation temperatures were chosen to be $T=0.1,0.2$ which corresponds to the region of hopping conduction        
in the case were the localization radius $a$ is of the order of 1 CU\cite{ourprl}.        
Decreasing the value of $a$ we may switch from the VRH to the nearest-neighbor hopping regime.       
The size of the system was taken to be $100 \times 100$ lattice sites.       
        
The simulation was performed using the following steps:
In the first stage the system was brought to ``thermal equilibrium'', inside a single PS, using the thermodynamic 
method employed in the work of Menashe {\it et al.}\cite{menashe}. In the second
stage, which is the reference point for the time from which       
the relaxation of the DS, energy and conductivity were studied,
the electron concentration was slightly changed by         
adding (or removing) electrons randomly. 
The time dependences were averaged       
over $10^3$ different sets of randomly distributed extra electrons (holes).       
Due to electron-hole symmetry of the Hamiltonian Eq.(\ref{eq:ham}) at $\nu=1/2$       
the relaxation of extra electrons and extra holes is the same. The data below        
describes the addition of electrons. It should be noted that
the insertion of electrons into the system can be 
simulated in many ways ,e.g., adding electrons
in a confined spatial region simulating the way electrons might enter the 
sample experimentally, or adding electrons to the most probable states. It is 
clear however that since the unperturbed system is saturated within one PS the 
difference between different electron insertion scenarios should only be 
notable within very short time scales i.e. of the order of our simulation times. 
Thus the choice of how to simulate the perturbation of the system by adding 
charge does not in any way effect the long time relaxation of the system which 
is due to transitions between different PS's.

\begin{figure}         
\includegraphics[width=8.6cm]{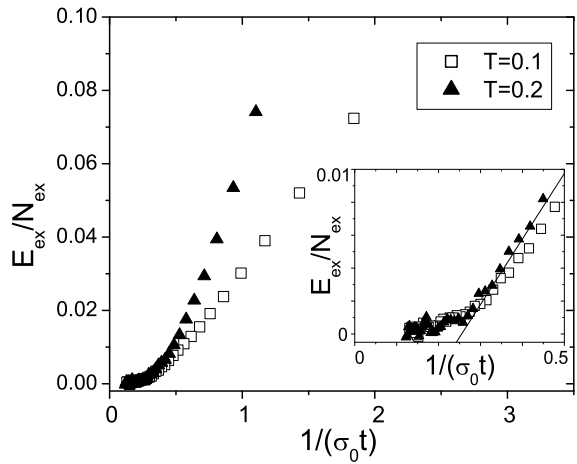} \vspace{0.5cm}         
\caption{The time dependence of the total energy of the system is shown at two different        
temperatures as a function of $1/\sigma_0 t$. The localization radius $a=1$,       
the system size is $100\times 100$.       
The inset shows the range of time where two curves coincide. The slope, that follows from        
Eq.(\ref{exact}) is shown by the straight line. All values are given in computer       
units. }         
\label{fig:1} 
\end{figure}       
        
\begin{figure}         
\includegraphics[width=8.6cm]{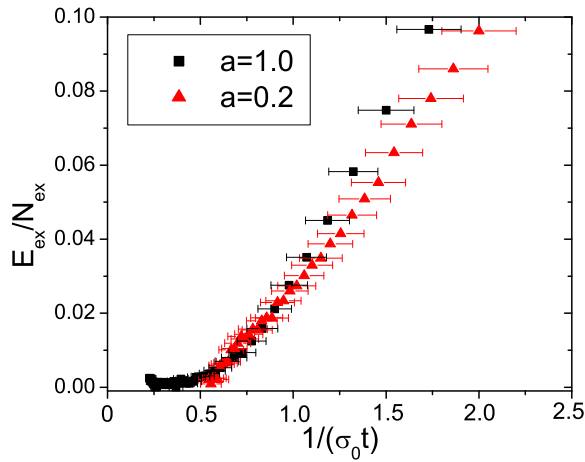} \vspace{0.5cm}         
\caption{The  energy relaxation in  the VRH       
($ a=1$) and in the  nearest neighbor hopping ($a=0.2$) regimes at $T=0.1$.}         
\label{fig:2}         
\end{figure}

\begin{figure}         
\includegraphics[width=8.6cm]{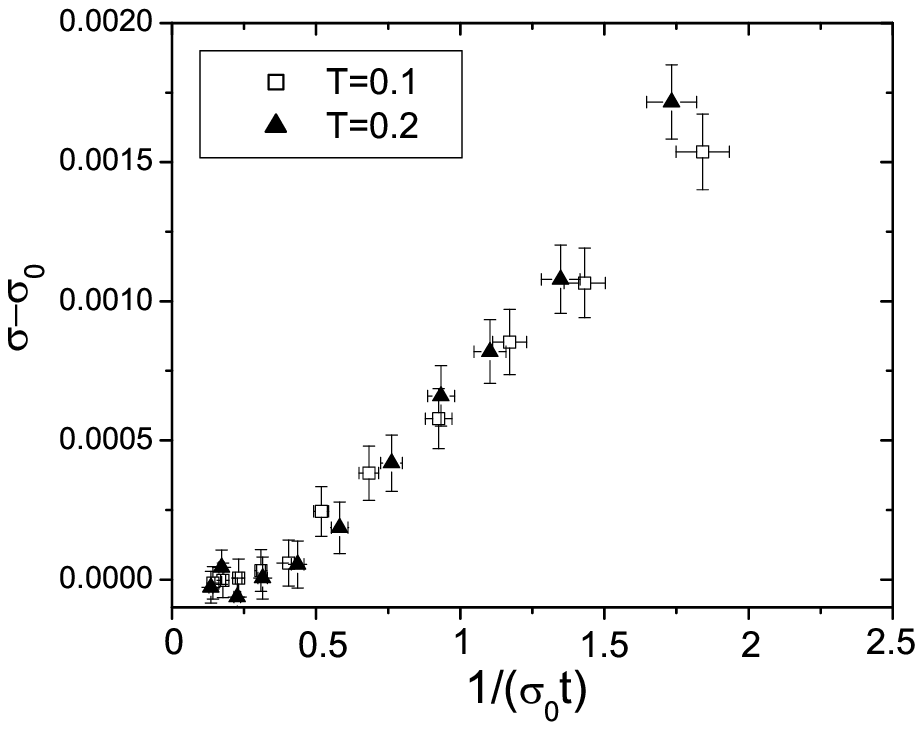} \vspace{0.5cm}         
\caption{The time dependence of the conductivity of the system is shown at different        
temperatures as a function of $1/\sigma_0 t$. The value of        
$\sigma_0 t$ is taken to be the saturated value of the conductivity at the longest        
time measured. The concentration        
changes at $t=0$ and the parameters of the simulation correspond to those of Fig.1.}        
\label{fig:3}         
\end{figure}

The energy relaxation is shown in  Fig.\ref{fig:1} and Fig.\ref{fig:2}. The reference point for 
the total  energy $E_{ex}$ is the saturated value for the longest        
time measured. At time $t=0$ the filling factor $\nu$ has been increased from 0.5 to 0.52, such that the total number of extra 
electrons equals $N_{ex}=0.02L^2/2 $. Fig.\ref{fig:1} shows the regime of the VRH with $a=1$ at two 
different temperatures. In Fig.\ref{fig:2} we 
compare relaxations of energy in  the VRH regime at $a=1$ with the relaxation in the nearest neighbor hopping at $a=0.2$.        
The values of the VRH conductivity at $a=1$ are $\sigma_0=0.0048,0.021$ in CU for $T=0.1,0.2$ respectively. These values of 
conductivity are also obtained as the saturated value at the longest time measured. Since Eq.(\ref{exact}) shows that at        
large $t$, the energy $E$ is a function of $\sigma_0 t$, we use this product       
as a reciprocal argument in Figs.\ref{fig:2},\ref{fig:3}.         
The straight line in the inset of Fig.\ref{fig:1} shows the slope as given by        
Eq.(\ref{exact}).       
The inset shows the final stage of the relaxation where the curves at        
different temperatures coincide and obey the time dependence given by        
Eq.(\ref{exact}). The saturation at even larger $t$ is due to a size effect.        
Note that the values of the conductivity $\sigma_0$ differ by almost five        
times for the two curves presented. The observed saturation of energy relaxation       
is connected to the finite size effect as discussed in a previous section.       
The relaxation of the charge fluctuations given by Eq.(\ref{exact})        
not only has the universal form for the different values of        
conductivities, but it is also independent on the conduction mechanism.
In the Fig.\ref{fig:2} we present the relaxation curves for the system       
for VRH conductivity and for nearest-neighbor hopping where the        
localization radius of the electrons is much smaller than the        
distance between neighboring sites. The time region where the two curves       
    coincide within our accuracy is even broader despite the        
fact that the values of the conductivity differs by a factor of $3000$.        
The main source of the errors (see Appendix) arises from calculating the values of the conductivity
$\sigma_0$ rather than measuring the total excessive energy of the system.       
              
Next we present the results for relaxation of the conductivity after        
adding extra electrons to the system. The procedure for adding the electrons
is the same  as stated above. The time dependence of the conductivity of the        
system is shown on Fig.\ref{fig:3} for two different temperatures. These were obtained        
in the same simulation as the data for the relaxation of the energy in Fig. \ref{fig:1}.        
One can see that for both temperatures the conductivity decreases with time finally        
reaching a kind of saturation, which may also be interpreted as a transition to a substantially slower        
rate of relaxation. We consider this 'saturation' as the end of the short time relaxation in our        
finite system. The value of the conductivity at the largest time we considered, is denoted by        
$\sigma_0$. One can see that      
as a function of $1/\sigma_0 t$ the results for both temperatures coincide even in an even wider time region 
than  for the total energy of the  system. The reason that the conductivity decreases is that, due to the non-equilibrium
the extra electrons occupy states above the        
Fermi level therefore providing higher current than in thermal equilibrium. 
Note that the saturation of conductivity occurs
 at  the same value of $\sigma_0 t$, that is $\approx 4$ as the saturation of energy. 
It is reasonable to think that the two processes are connected and the saturation of the conductivity is also a size effect.        
       
The time corresponding to the saturation $t=4/\sigma_0$ is very short. For T=0.1       
and $a=1$ it is $\approx 0.8 ns$. Even if we assume an infinite system and extrapolate the law 
$\sigma -\sigma_0 \sim (\sigma_0 t)^{-1}$ to much larger times, we find that in the microsecond range  
all relaxation which can be observed will be over.       
       
The saturation of the conductivity averaged over different initial       
distributions of extra electrons, which we have presented in this section, happens       
at very short times. It is a Maxwell-type relaxation of the extra charge. We have checked that the relaxation of the 
Coulomb gap inside the energy interval that is responsible for the VRH occurs even faster than the relaxations of the 
energy and conductivity. These results do not support the idea that long time
relaxation is due to the slow formation of the Coulomb gap\cite{yu}.        
      
However in the experimental data \cite{zvi1},\cite{zvi2} the long time relaxation       
happens on time scales of the order of seconds
and hours. These time scales are 9-11 orders of        
magnitude longer than those which we are able to simulate. Thus, if a long time relaxation       
exists in this system, the change of the averaged value of the conductivity is        
negligible during the physical time scales we have simulated. This explains the
apparent saturation of the conductivity, we observe.  If the long time 
relaxation results from the transitions between different PS's, the conductivity we have observed should be considered as
the conductivity within  a single PS. The encouraging result       
obtained in this section is that the saturation of the conductivity of a single PS can be achieved 
during the time scale available for our computation.      
Based on this result another approach to the problem of long time relaxation,        
is proposed in the next sections.       
       
\section{Conductivity of different pseudoground states in the lattice model}
\label{sec:conductivity}      
The idea behind our new approach is rather simple.  We want to compare the conductivity of the       
system in the different PS's. If the values of the obtained conductivities are different for       
different PS's then the long time relaxation of the conductivity can be attributed to the
slow transition between those PS's. We are unable to measure        
time scales of the order of such transitions still we can explain its effect and magnitude which       
are reflected through the difference in the conductivities of different PS's. In the experiments of        
the group of Ovadyahu \cite{zvi2} such differences are of the order of 10\%.       
       
To observe the conductivity of different PS's, we measured the conductivity of the same sample with the different 
initial distributions of electrons during the longest time we are able to simulate. The sample is characterized by 
the total set of random energies $\phi_i$. Starting with different initial distribution of electrons        
the system relaxes to the different PS's. If the saturated values of
conductivities, measured as in previous section, is different, one should expect 
that the system will have a long time relaxation 
due to transitions between different pseudoground states.       

\begin{figure}         
\includegraphics[width=8.6cm]{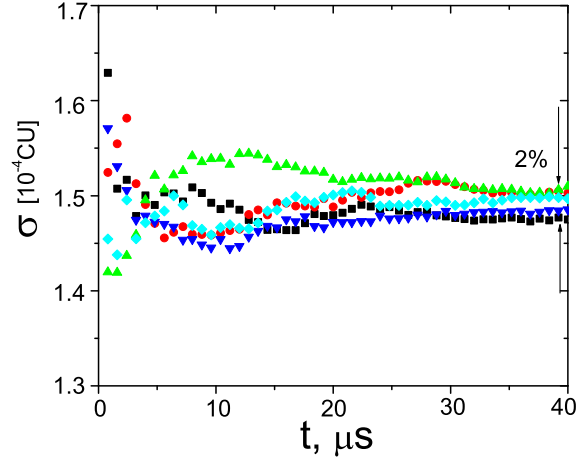} \vspace{0.5cm}         
\caption{The time evolution of the conductivity averaged over the time        
of measurement is shown for different initial distributions of electrons  in the lattice model for the same 
set of $\phi_i$. The values of parameters are $A=1$, $T=0.04$,        
the localization radius $a=1$, the system size $L=70$, and the filling factor $\nu =1/2$.}         
\label{fig:4}         
\end{figure}

The results for the lattice model simulations performed at the lowest temperature are         
presented in  Fig.\ref{fig:4}. The time evolution of the conductivity averaged over the time        
of measurement is shown for different initial distributions of electrons for the same       
sample. As one can see there is no appreciable difference in the values of       
the conductivities within our accuracy which is about $1-2\%$. We obtained the same result for       
all higher temperatures even with greater accuracy.       

We therefore conclude that there is no apparent difference in in the conductivities of different PS's
for this model down to the lowest temperatures we are able to simulate. We think that the reason for this is 
the following: It has been shown\cite{ourprl} that the VRH conductivity of the system       
is provided by the single-electron excitations. The properties of these excitations is 
defined by the structure of the Coulomb gap. In the lattice         
model at large $A$ the  DS in the Coulomb gap has a universal form\cite{pik}. It is independent       
of the properties of the system and is the same for all PS's. For our
temperature range, the case  $A=1$ can
be considered as a large disorder. Thus all PS's have the same conductivities        
and no long time relaxation can be observed.
We  realize that the lattice model with large disorder can not  account for
the effect of the long time relaxation of the conductivity which is observed experimentally. Thus,        
to observe the difference in the conductivities of different        
PS one should take a system with smaller disorder, where the Coulomb gap is not        
universal. Unfortunately, it is difficult to find such a regime in the lattice model,        
due to the Wigner crystallization at low temperatures.

\section{Conductivity of different pseudo ground states in the random site model} 
\label{sec:randomsite}      
In this section we present the results for the random site model.  
The Hamiltonian of the model has the form
    
\begin{equation}        
H = \frac{1}{2} \sum_{i \neq j}        
\frac{\left(n_i - 1/2 \right) \left(n_j - 1/2 \right)}{r_{ij}}.        
\label{eq:ham1}        
\end{equation}  
It differs from Eq(\ref{eq:ham}) in two important respects. The first: it is formulated on sites i,j, 
which have random positions on the plane. The second: any random energies $\phi_i$, that are not 
correlated with the interaction, are absent. Thus, the random
positions of the sites is the only source of disorder. 

We consider the case $\nu=1/2$ for which each state of the system has an exact twofold degeneracy: the total energy is 
invariant with respect to changing all of the occupation numbers $n_i\rightarrow 1-n_i$. It is this 
symmetry for half filling which allows this model to be mapped on to a spin glass model\cite{Grannan}.
Probably, this fact is also important for formation of PS's that consist of fragments of both states.

This model is unusual for electronic systems. For example partially occupied donors located in a plane with a gate 
electrode above the plane are not described by this model. In this case the above symmetry is absent. However, the neutral 
system of random donors with a large ``negative U'' can be described by the Hamiltonian Eq.(\ref{eq:ham1}), 
if 1/2 of them have double occupation with charge -1 and the other half are not occupied with the charge 1.
     
For the simulation of the random site model we use a similar computational algorithm which is described in the Appendix.
Unfortunately, it is more time and memory consuming than the algorithm for the lattice model. In this model the unit of length  
is $a_0=n^{-1/2}$, where $n$ is the  concentration of sites.       
       
\begin{figure}         
\includegraphics[width=8.6cm]{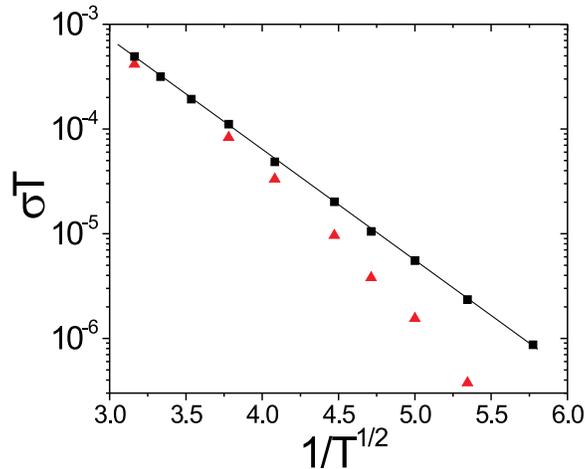} \vspace{0.5cm}          
\caption{The temperature dependence of the VRH conductivity in the random        
site model (triangles) and in the lattice model (squares) is shown. The disorder       
strength is given by $A=1$ for the lattice model, while $A=0$ for the random site       
model. The localization radius $a=1$ and the filling factor $\nu =1/2$. The straight        
line represents ES law.}         
\label{fig:5}         
\end{figure}

Fig.\ref{fig:5} shows the temperature dependences of the VRH conductivities both in the random        
site model and in the lattice model. Unfortunately for the random site model we were unable to check the importance of
simultaneous many-electron transitions on the VRH conductivity.      
One can see a deviation from the ES law in the case of the random site model,
that might be  a result of deviation of the DS from the universal DS in the Coulomb gap.        
       
\begin{figure}         
\includegraphics[width=8.6cm]{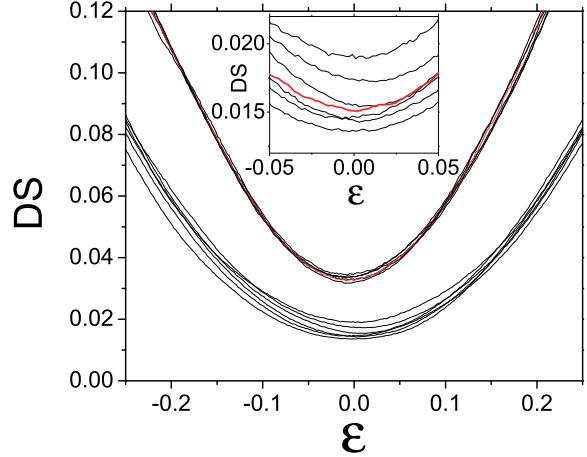} \vspace{0.5cm}         
\caption{The DS in the vicinity of the Fermi level at $T=0.04$ as a function of the single-particle energy 
$\epsilon$ with a reference point at the Fermi level is shown for the lattice model (the upper  set of        
curves) and random site model (lower curves) is presented for $5$ different initial        
distribution of electrons for  each model. The DS is averaged over a time equal to       
$1\mu s$. The inset shows the DS, near the Fermi level, for the random site model in an enlarged scale.
The thermodynamic DS is emphasized in the inset by a thick line. The values for the parameters are the same as were employed
in Fig.\ref{fig:5}, except that $A=0$ is used for the random site model.}        
\label{fig:6}         
\end{figure}

Fig.\ref{fig:6} confirms this point of view. It shows DS of different PS's for both       
lattice and random site models. As in the previous section the different PS's       
have been obtained by simulations starting from different initial distribution of electrons but with 
the same disorder. The latter condition means the same set of $\phi_i$ or the same set       
of random sites depending on the model. In each case the DS is calculated by       
two methods. The first is the kinetic Monte-Carlo, which gives the DS in       
one PS due to the time constraint, and the second is the thermodynamic Monte-Carlo method which gives       
the DS averaged over all PS's. To get to the thermodynamic regime one should ignore        
the tunneling exponent in the transition rate so that transition at any distance becomes possible. However 
the energy dependence of the rate should be strictly preserved to get the correct ergodic thermodynamic result. 
Menashe {\it et al.}\cite{menashe} have shown that this method provides an effective thermalization 
including fast transitions between the PS's. 
       
Fig.\ref{fig:6} contains two important results:
(i) The DS of the random site model        
strongly differs from the DS of the lattice model and it does not have  the standard        
energy dependence typical for the two-dimensional Coulomb gap, which is known        
to be very robust in the lattice model for A=1. For the random site model the DS is quadratic rather than linear. 
We have checked that its curvature is $T$-independent. 
From dimensionality considerations this DS can only be of the order of $ \epsilon ^2/e^6 n^{1/2}$, which also      
explains the deviation from the ES law for the VRH conductivity shown in Fig.\ref{fig:5}. 
(ii) The relative difference between       
the DS for different PS is much larger for the random site model than for the lattice model. For the random site model 
the time fluctuations of DS near the Fermi level        
are $10$ times smaller than the difference between the  DS's of different PS's. In the lattice model  those fluctuations  are
so close to the difference itself that the difference is not a reliable measure. 
The thermodynamic DS is between the DS's obtained for the different PS's as it should be for an average function.   

In fact results (i) and (ii) are connected to each other. At large $A$ the universal behavior 
of the Coulomb gap can be obtained from conditions $\epsilon_j-\epsilon_i -1/r_{ij}>0$ for every 
empty site $j$ and occupied site $i$. These conditions are important near the Fermi energy only, for which 
the total energy becomes irrelevant. That is the reason why all PS have similar DS near the Fermi level. 
For the random site model these conditions are also necessary but there should be some other 
restrictions that make the DS smaller. Those restriction should be connected with the average 
distance between the sites and therefore they are related to the total energy. 
Therefore it is interesting to study the conductivity of different PS in this model.        
       
\begin{figure}         
\includegraphics[width=8.6cm]{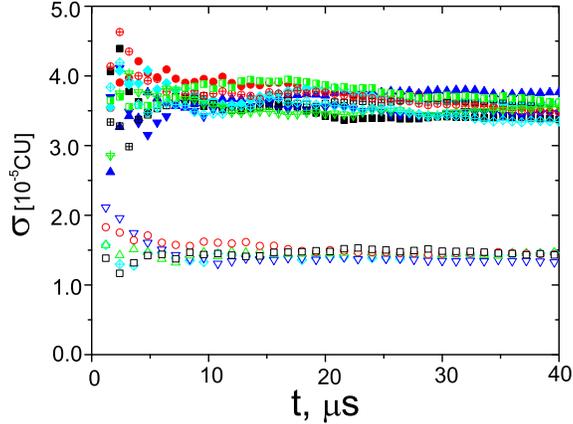} \vspace{0.5cm}         
\caption{The time evolution of the conductivities of different PS's       
for the model with the random  site distribution are shown by filled symbols.
Open symbols show the same evolution for a ``combined'' system with A=1. The
values of parameters: $T=0.04$, localization radius $a=1$, the system size $L=70$      
and the filling factor $\nu =1/2$.}         
\label{fig:7}         
\end{figure}       
       
The simulation results for the time evolution of the conductivities of different PS's       
for the model with the random spatial site distribution are shown in Fig.\ref{fig:7}
by filled symbols  at the same temperature as for the lattice model. In this case         
the conductivity value of each individual PS saturates within 1\% during the time        
of the simulation. However, the saturated values of the conductivities       
differ by 12\% for the different initial distributions of electrons in the same sample.       
This is by an order of magnitude greater than the simulation accuracy.       
The study of the conductivities in  the random site model shows        
that the slow transitions between PS's may result in the long time relaxation of the       
conductivity observed in the experiment.
An important question now is whether the obtained result is due to the randomness in the 
site distribution or to the absence of the disorder which is not correlated with the interaction. 
       
In order to answer this question  we consider a       
``compound'' model. Namely we have added to the Hamiltonian of the random site model
Eq.(\ref{eq:ham1}) the first term in Eq.(\ref{eq:ham}) with $A=1$. 
One can see from Fig.\ref{fig:7} that the difference in the conductivities of different PS's 
disappears at $A=1$ within the  simulation accuracy. Thus, the randomness of the sites is not important, 
but the value of $A$ is very important.     
\begin{figure}         
\includegraphics[width=8.6cm]{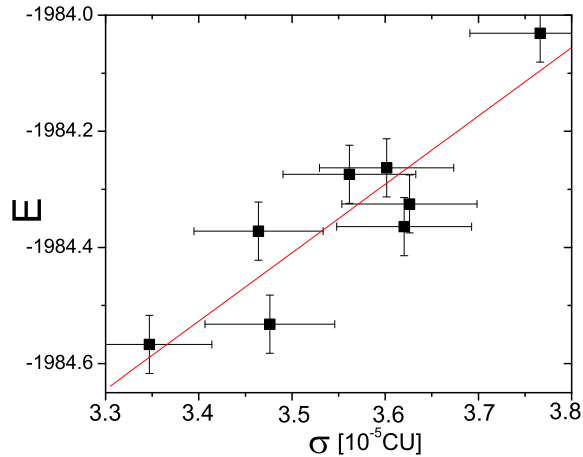} \vspace{0.5cm}         
\caption{The correlation between the total energy of the system and the conductivity        
is shown for 8 different initial distributions of electrons in the random site model.        
Both the total energy and the conductivity are averaged over the time        
$40\mu s$. The straight line is given as a guide for the eye. The values of the parameters used are the 
same as were used in Fig.\ref{fig:7}.}         
\label{fig:8}         
\end{figure}       
          
Another important question is whether the values of the conductivity of the        
different  PS's are correlated with the energies of these PS's. The total energy of the system in a given PS is averaged over time       
and plotted versus the value of the conductivity for the same PS's in Fig. \ref{fig:8}.           
The data  shows that the energy dependence of the conductivity is close to a linear behavior. PS's with higher average 
total energy have also a larger value  for the conductivity which is reasonable, since in the states with lower energy        
the electrons are in positions where they are more tightly bound and therefore their conductivity is lower. 
The same behavior is apparent in the experimental results of Ovadyahu      
\cite{zvi2}. The linear dependence is also reasonable because the energy  difference between  PS's is 
small and conductivity as a function of energy may be expended into a Taylors series, taking the first term. Similar to 
Anderson {\it et al.}\cite{anderson}
one can show that possibility of a Taylor's expansion leads to the relaxation
law $\delta\sigma \sim \log t$.          

\section{Conclusions}       
       
A novel computational algorithm is presented which permits one to simulate the energy,        
density of states and conductivity of a system with localized interacting        
electrons during times of the order of 40$\mu$s. We argue that during this time the relaxation of the system 
to some pseudoground state is completed. An analytical theory of the energy relaxation  which 
is in a good agreement with the computational data is presented.       
Our computational results for the conductivity exhibit two very distinct time scales: the first is
a very short time scale corresponding to the average value of the conductivity, the second a very long time 
scale defined by the long time relaxation of the conductivity. We attribute these two scales to 
the following physical picture, in which the relatively short time scale is a consequence of the 
relaxation of the system within one PS, well described by our analytical theory, as well as by our simulations,
whereas the long time relaxation of the conductivity, is related to transitions between different       
pseudoground states. The microscopic origin of this huge time scale separation can be attributed to the fact
that whereas many-electron transitions are not important for the VRH conductivity within one PS they
play an important role in slowing down transitions between PS. 

Current computational resources are not able to confirm this theory by directly observing 
the long relaxation processes, rather they are limited to the range of the short relaxation time scales,
which can be simulated. In order to check our theory we have studied the conductivities of the different 
pseudoground states to see whether or not they are different. We employed two different models:
the lattice model and the random site model, with random sites and no external disorder. We have shown that
these two different models which correspond to different realizations of disorder lead to different physical effects.
For the lattice model no difference in the conductivities of the different 
pseudoground states has been found. We understand this result in terms of universality of the Coulomb gap 
and ES hopping conduction.
           
For the model with random distances between sites and no external disorder we have found the difference of the 
conductivities to be within 10-12\% which is large enough to explain the experimental data. 
We have shown also that the density of states in this model is not universal and that hopping conductivity does not 
obey the ES law. We think that similar effects might be observed in the lattice model with
$A=1$ as well as at lower temperatures than those which we are able to achieve. 
With increasing $A$ this temperature range should become lower. 
Thus we think that universality of the Coulomb gap, that manifests itself in the        
ES law for the VRH, suppresses the long time relaxation because in this case the       
conductivities of different pseudoground states are very close to each other.         
The work has  been funded  by the US-Israel Binational        
Science Foundation  Grant 9800097.  The computations have been made in CHPC of        
the University of Utah. A.E. is grateful to Aspen Center for Physics, where this paper has been presented for the first time, 
 and to Boris Shklovskii and Zvi Ovadyahu for important comments.          
       
\section{Appendix. Computational algorithm.}       
        
To perform the simulation of the transport and thermodynamic 
properties on the finite array $L\times L$        
for both the lattice model and the model with random spatial site distribution       
we use the periodic boundary conditions on a torus.        
In fact this means, that for the pair of sites $i$        
and $j$ the distance between them is given by         
$r_{ij}=[(\Delta x_{ij})^2+(\Delta y_{ij})^2]^{1/2}$, where
\begin{equation}         
\Delta x_{ij}=min\left(|x_i-x_j|,L-|x_i-x_j|\right)\\
\Delta y_{ij}=min\left(|y_i-y_j|,L-|y_i-y_j|\right).
\label{deltaxy}
\end{equation}        
Here ${x_i}$ and $y_i$ are the sets of the site coordinates, which form
 a lattice in one model and are random in the other.       
        
To simulate the conductivity one should add a term $\sum_i Ex_i$ to the 
Hamiltonian Eq.(\ref{eq:ham}) where $E$ is a weak electric field.
Due to the field the current flows around the torus in $x$ direction.
It is convenient to calculate the total dipole moment due to electron
transitions in the direction of electric field and obtain the conductivity
from the equation
      
\begin{equation}        
\sigma=\frac{1}{EL^2}\frac{dP}{dt},        
\label{cond}        
\end{equation}        
On average the dipole moment $P$  increases linearly  with time.
        
To find $dP/dt$ one needs a kinetic Monte Carlo(MC) algorithm, that connects
the number of MC steps with a real time $t$. Note, that for any 
thermodynamic calculations the time is irrelevant. The starting equation
is a transition rate for a single electron hop from site $i$ to site $j$,
that has a form
     
\begin{equation}       
     \Gamma_{ij}=\gamma \theta_{ij} \frac{\exp(-2r_{ij}/a)}{1+\exp(\epsilon_{ij}/T)},        
\label{rates}        
\end{equation}        
where  $r_{ij}$ is the distance between the sites $i$ and $j$,         
$\epsilon_{ij}$ is the energy difference between the two configurations         
$\epsilon_{ij}=\epsilon_{i}-\epsilon_{j}-1/r_{ij}-E\Delta x_{ij}$,        
$\epsilon_{i}=\phi_{i}+\sum_{j} {1 \over {r_{ij}}} \left(n_j - \nu \right)$,
$\Delta x_{ij}$ is given by Eq.(\ref{deltaxy})
and $\theta_{ij}$ is equal to $1$ if the site $i$ is occupied and site $j$ 
is empty or $0$ otherwise.
The transition rate should have the dimensionality of frequency, it is written
in a dimensionless form, assuming  our time unit is $\gamma^{-1}$. The MC process
can be started with any initial set distribution of the occupation numbers $n_i$ which
evolves during the simulation.

There are two different algorithms developed for this type of computer        
simulations. The first one implies the calculation of all the transition        
rates $\Gamma_{ij}$ in the system at each Monte Carlo (MC)       
 step. Then the probability that the next transition to occur is        
$i \rightarrow j$ is given by       
\begin{equation}       
\frac{\Gamma_{ij}}{\sum_{i}\sum_{j\neq i}\Gamma_{ij}}.         
\end{equation}        
Then at each MC step the code chooses the transition with the above probability       
and performs it. This means changing the occupation numbers         
$n_i$ and $n_j$, calculating the contribution of this transition into the        
total dipole moment $P$ and       
recalculating all site energies and transition rates.       
After that the code comes to the next MC step.       
                
 For the above algorithm the physical time $\Delta t$ for each MC step is 
\begin{equation}
\Delta t=(\sum_{i} \sum_{j \neq j}\Gamma_{ij})^{-1},
\label{deltat1}        
\end{equation}
because in the real system all processes run simultaneously. 
Note, that $\Delta t$ depends on the configuration of the system and varies during the simulation.       

In the second algorithm at each MC step a pair of sites $(i,j)$ is chosen 
with equal probability from all possible       
sets. Then the transition $(i \rightarrow j)$       
is accepted with the probability $\Gamma_{ij}$. 
If the transition is rejected then the MC step is over and code        
proceeds to the next MC step. If it is accepted the transition 
is performed.
This means that code changes occupation numbers $n_i$ and $n_j$,
calculates the contribution of this transition into the
total dipole moment $P$, and  recalculate all the site energies
$\epsilon_{ij}$. This is the end of the MC step.
In this case, the physical time per  one MC step is constant and equal to
\begin{equation}
\Delta t=\frac{1}{N_{tr}}=\frac{1}{L^2(L^2-1)},
\label{deltat2}        
\end{equation}         
 $\Delta t=1/N_{tr}$, 
where $N_{tr}$ is the total number of different transitions in the system.          
       
The advantage of the first algorithm is that any MC step is successful:       
as a result of each step one electron moves from one  site       
to another. 
The disadvantage is that for the interacting systems at each MC step        
the computer should recalculate $N_{tr}$ values of $\Gamma_{ij}$.       
Therefore each MC step is very time consuming as compared to the MC step       
of the second algorithm. However,        
the disadvantage of the second algorithm is that at strong dispersion         
of the transition rates (small $a$ or $T$) the probability of the rejection 
is very high. In other words, the physical time $\Delta t$ per        
one MC step is much smaller than in the first algorithm. This can be seen
from Eqs. (\ref{deltat1},\ref{deltat2}). Indeed, the double sum in Eq. 
(\ref{deltat1}) contains $N_{tr}$ terms. However, the majority of these
terms are very small. 

In this paper, we used a mixed scheme which combines the best        
features of both algorithms discussed above. We show that it is very        
efficient in the VRH regime for the interacting electrons. The original idea for        
this algorithm belongs to O.Biham \cite{ofer}.       
       
The transition rate Eq. (\ref{rates}) for the VRH can be written as a product
$\Gamma^T_{ij} \Gamma^A_{ij}$, where the        
'tunneling' part of the transition rate        
$\Gamma^T_{ij}= \exp(-2r_{ij}/a)$,        
while $\Gamma^A_{ij}=\theta_{ij}/(1+\exp(\epsilon_{ij}/T))$       
reflects activation. It is important now that $\Gamma^T$ is independent of
the configurations of electrons and should be calculated only once. Since
the probabilities of tunneling and activation are independent we may apply
the first algorithm with $\Gamma^T$ and the second one with $\Gamma^A$.       
Practically it means that we choose a pairs $(i,j)$ with the
probabilities       
\begin{equation}       
\frac{\Gamma^T_{ij}}{\sum_{i}\sum_{j \neq i}\Gamma^T_{ij}}       
\label{weights}       
\end{equation}       
and accept the transition $i \rightarrow j$ with the probability       
$\theta_{ij}/(1+\exp(\epsilon_{ij}/T))$.        
If the transition is rejected then the MC step is over and the code proceeds 
to the next MC step. If it is accepted, the transition is performed.       
To finish this step the code changes the occupation numbers $n_i$ and $n_j$,
calculates the contribution of this transition into the        
total dipole moment $P$, and recalculate all the site energies       
 $\epsilon_{i}$.       
In this case,        
the physical time per  one MC step is constant and equal to
\begin{equation}   
\Delta t=(\sum_{i} \sum_{j \neq i}\Gamma^T_{ij})^{-1}.       
\label{deltat3}
\end{equation}         
For the lattice model the sum $\sum_{j \neq i}\Gamma^T_{ij}$ is independent on $i$ thus        
$\Delta t=(L^2 \sum_{j \neq i}\Gamma^T_{ij})^{-1}$.       
Using this result  the conductivity in units of $\gamma a_0$ can be written in a form       
       
\begin{equation}        
\sigma=\frac{1}{EL^2}\frac{dP}{dt}=\frac{P\sum_{j\neq i} \exp{(-2r_{ij}/a)}}{EN_{MC}},        
\label{condMC}        
\end{equation}        
where $P$ is the total dipole moment due to the electron transitions after         
$N_{MC}$ steps.        
       
For the model with random spatial site distribution the $\sum_{j \neq i}\Gamma^T_{ij}$       
is different for each site $i$ and the conductivity is given by       
\begin{equation}        
\sigma=\frac{P\sum_i \sum_{j\neq i} \exp{(-2r_{ij}/a)}}{EN_{MC}L^2}.        
\label{condMCrand}        
\end{equation}             
       
\begin{table}       
\caption{The algorithm efficiency comparison.} 
\scalebox{0.7}{\includegraphics{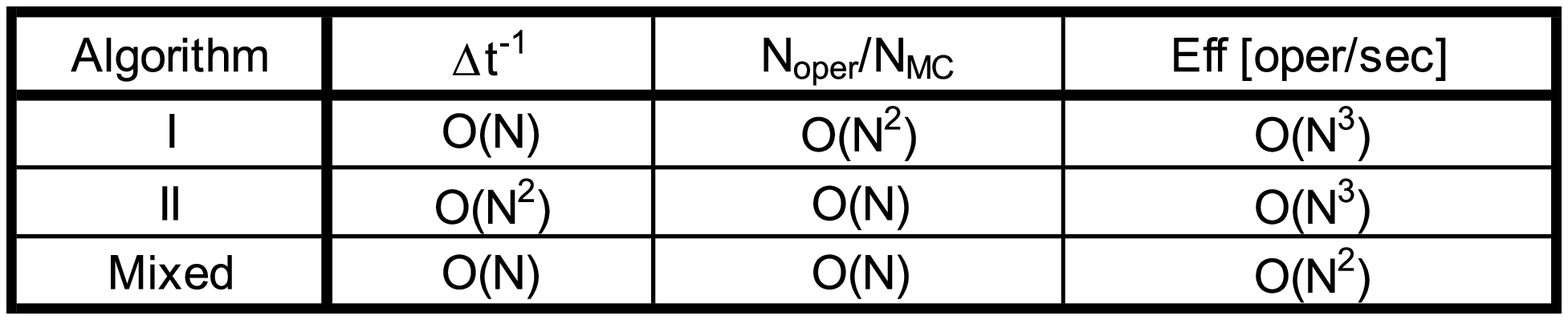}}        
\label{algorithmtable}         
\end{table}       
       
Now we compare the efficiency of all three algorithms. The efficiency       
is the number of operations which are necessary to simulate       
a physical process during a time $t$. The most important is how the efficiency depends  on the number
of sites in the system $N=L^2$. The results are shown in the Table 
\ref{algorithmtable}. In the table $\Delta t$ is the physical time 
correspondent to one MC step, $N_{OP}/N_{MC}$       
is the number of operations per MC step, and the efficiency is given by        
$Eff=N_{OP}/N_{MC}\Delta t$. 

The time efficiency of the mixed algorithm is the same for both lattice and
random site models. However, the memory        
requirements are much harder for the random site model. In this model one need to        
calculate all $N^2=L^4$ tunneling terms and have access to all of them at each MC step,       
because the transition at each step is chosen with the above weights. While in the        
lattice model there is only $N=L^2$ different tunneling terms $\exp{(-2r_{ij}/a)}$ that       
have to be stored. In fact, this constraint does not allow us to simulate a system in which       
the number of sites exceeds 5000 employing the random site model.       

One can see from the Table \ref{algorithmtable} that the efficiency of the 
mixed algorithm is the best. The use of this algorithm allowed us to simulate 
the macroscopic conductivity. In fact this algorithm has been used
in Ref\cite{ourprl}.        
       

        
\end{document}